\documentclass[%
 reprint,
 amsmath,amssymb,
 aps,
]{revtex4-1}

\usepackage{graphicx}% Include figure files
\usepackage{dcolumn}% Align table columns on decimal point
\usepackage{bm}% bold math
\usepackage{natbib}
\usepackage{graphicx}
\usepackage{breqn}
\usepackage{siunitx}
\usepackage{gensymb}
\usepackage{xcolor}
\usepackage{caption}
\usepackage{subcaption}

\usepackage[normalem]{ulem}

\begin{document}

\preprint{APS/123-QED}

\title{Orientation gradients in rapidly solidified pure aluminum thin films: comparison of experiments and phase-field crystal simulations}% Force line breaks with \\

\author{Paul Jreidini$^1$}
\author{Tatu Pinomaa$^2$}
\author{Jörg M.K. Wiezorek$^3$}
\author{Joseph T. McKeown$^4$}
\author{Anssi Laukkanen$^2$}
\author{Nikolas Provatas$^1$}
\affiliation{$^1$Department of Physics and Centre for the Physics of Materials$,$ McGill University$,$ Montreal$,$ Canada \\
$^2$Integrated Computational Materials Engineering group$,$ VTT Technical Research Centre of Finland Ltd$,$ Finland\\
$^3$Department of Mechanical Engineering and Materials Science$,$ University of Pittsburgh$,$ USA\\
$^4$Materials Sciences Division$,$ Lawrence Livermore National Laboratory$,$ USA
}

\date{\today}

\begin{abstract}

Rapid solidification experiments on thin film aluminum samples reveal the presence of lattice orientation gradients within crystallizing grains. To study this phenomenon, a single-component phase-field crystal (PFC) model that captures the properties of solid, liquid, and vapor phases is proposed to model pure aluminium quantitatively. A coarse-grained amplitude representation of this model is used to simulate solidification in samples approaching micrometer scales. The simulations reproduce the experimentally observed orientation gradients within crystallizing grains when grown at experimentally relevant rapid quenches. We propose a causal connection between defect formation and orientation gradients.

\end{abstract}

\maketitle

Several modern industrial processes, such as thermal spray coating deposition \cite{lavernia2010}, certain welding techniques \cite{david2003}, and metal additive manufacturing \cite{debroy2018}, operate under rapid solidification conditions. Rapid solidification leads to drastically altered microstructures through selection of metastable phases, solute trapping kinetics \cite{cahn1998,aziz1996,pinomaa2020MRS}, changes in solidification morphology  \cite{trivedi1994solidification}, and reduction of dendrite primary arm spacing \cite{trivedi1994solidification}. An overlooked key feature of these microstructures is the formation of various types of crystalline defects \cite{balluffi2005}. These include trapping of excess point vacancies \cite{zhang2017}, formation of high dislocation densities \cite{wang2018,wang2020}, high microstructural (type II-III) residual stresses \cite{chen2019}, and so-called lattice orientation gradients \cite{polonsky2020}.
As these crystalline defects critically affect materials' mechanical properties and performance, improving our understanding of their formation mechanisms leads to significant advances in the aforementioned processing techniques.

Among the above listed classes of crystalline defects, lattice orientation gradients formed during solidification remain poorly understood. These defects consist of a gradual change in crystallographic orientation within a single grain, in contrast to the usual concept of an interface between two misoriented grains across which the orientation changes abruptly. Their observation is relatively novel in rapidly solidified samples, and we are not aware of evidence for orientation gradients in slower solidification regimes.
This work aims to remedy the resulting dearth of knowledge on the origins and characteristics of such defects. To that end, we first report on orientation gradients and related defect structures recently observed in rapidly solidified experimental samples of pure aluminum. We then develop a theoretical model capable of producing comparable observations, and use simulation results to rationalize the orientation gradients' formation mechanisms.

\begin{figure*}
\centering
\includegraphics[width=\textwidth]{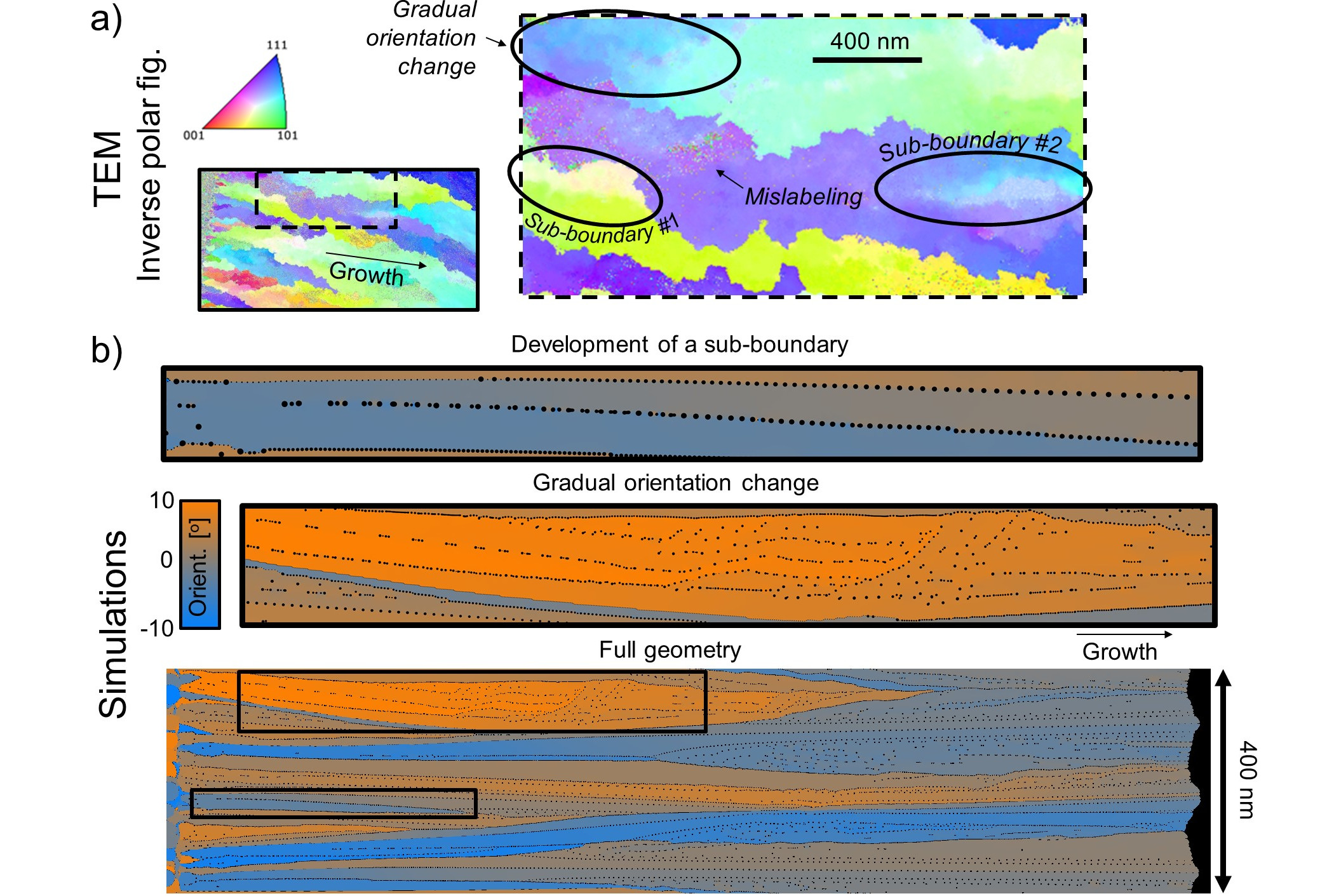}
\caption{
Orientation distribution in rapidly solidified aluminum thin film, for a) a laser thin film experiment as an inverse pole figure map, and b) in a phase-field crystal simulation.
The color in a) represents the crystallographic plane oriented in the plane of the image. 
The color in b) represents a two-dimensional lattice orientation in the film plane, except for black, which represents non-crystalline areas such as liquid, voids or dislocations. Orientation gradients are clearly visible within individual grains in a) and b). The overall growth directions are indicated with thin black arrows. Top frames in b) are zoom-ins of the insets in the bottom frame. 
}

\label{fig_orientationGradients}
\end{figure*}

Establishing a causal link between solidification conditions and the resulting defected microstructures requires analyzing
highly transient processes over small observational time and length scales, making the experimental investigation of rapid solidification structures challenging. 
Nevertheless, recent thin film resolidification experiments have provided high quality information about emergent rapid solidification microstructures \cite{zhong2010,kulovits2011,mckeown2016,zweiacker2018,bathula2020}. The general approach is to use a pulsed laser beam to create an elliptical melt pool in a thin polycrystalline sample, and then let the sample resolidify \cite{mckeown2016,zweiacker2018}.
We applied this procedure on a 100 nm thick polycrystalline sample of pure aluminium with a 50 nm thick amorphous Si$_3$N$_4$ underlay.
The as-solidified microstructure was imaged with a scanning/transmission electron microscope (S/TEM) on a spatial resolution of 5 nm and an angular resolution of 1$^o$. 
The resulting inverse pole figure map is shown in Figure \ref{fig_orientationGradients}a, where strong orientation gradients within grains are clearly visible and are highlighted with ellipses. These include a gradual change in orientation, and `sub-boundaries' which run approximately parallel to the growth direction and separate misoriented regions inside a single grain.
The smooth orientation gradients inside grains are often accompanied by pixels in the image data where the local orientation varies abruptly, indicated as `Mislabeling' in Figure \ref{fig_orientationGradients}a. A potential explanation for the mislabeling is a high concentration of crystalline defects. This hypothesis is explored later in this work.

Since the experimental techniques lack the spatial resolution and temporal tracking needed to observe the formation of the orientation gradients in-situ at the nanometer scale, a numerical approach is required to complement the investigation of this phenomenon.
This paper proposes the use of a phase-field crystal (PFC) model \citep{elder2007}, which can be considered an offshoot of classical density functional theory. Unlike traditional phase-field methods\citep{boettinger2002,pinomaa2020_DTEM}, PFC is capable of self-consistently retaining lattice orientation and other atomic information while evolving microstructure dynamically on diffusive time scales. We base the PFC model used in this paper on that proposed by Kocher et al. \citep{kocher2025}, which accounts for both liquid and vapor disordered phases as well as an ordered solid phase in a system consisting of a single atomic species. The existence of a vapor phase in the model is crucial under the assumption that extreme interface velocities in a pure material system would be capable of trapping vapor pockets, or `voids', in a manner analogous to solute trapping in alloy systems \citep{pinomaa2020}. For an atomic density field $n(\mathbf{x})$, our model's free energy functional is written as
\begin{equation}\label{eq_freeEnergy}
\begin{split}
F[n] =& \int d\mathbf{x} \left( \sum_{l=2}^4\frac{1}{l}p_l n(\mathbf{x})^{l}  + \sum_{l=2}^4\frac{1}{l}q_l n(\mathbf{x})n_{mf}(\mathbf{x})^{l-1} \right) \\
&- \frac{1}{2}\iint d\mathbf{x}_1 d\mathbf{x}_2 C^{(2)}(\mathbf{x}_1-\mathbf{x}_2)n(\mathbf{x_1})n(\mathbf{x_2}),
\end{split}
\end{equation}
where $n_{mf}(\mathbf{x})$ represents a Gaussian smoothing operation applied to the microscopic density $n(\mathbf{x})$, parameters $p_l$ and $q_l$ are temperature-dependant and set the relative energies of the three accessible phases in thermodynamic equilibrium, and $C^{(2)}$ is a two-point correlation function that determines the lattice structure of the ordered solid phase. Here, we choose $C^{(2)}$ to be the minimal correlation function that leads to two-dimensional triangular lattice structure, represented in Fourier space by a single Gaussian peak\citep{greenwood2010,oforiopoku2013}. While this PFC model could be simulated in three dimensions, practical computational constraints limit us to two. This limit does not significantly impact the comparison with the rapid solidification experiments, as those were done in thin-film geometry with a thickness much smaller than length and width.

\begin{figure}[h!]
\centering
\includegraphics[width=.5\textwidth]{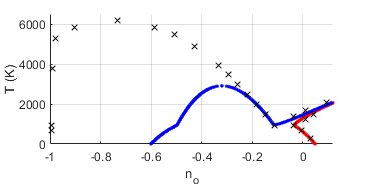}
\caption{Black crosses: thermodynamic phase diagram of pure aluminum\citep{lomonosov2007}. Blue: Coexistence lines of the disordered phases (liquid and vapor) of the proposed PFC model. Red: Coexistence line of the ordered (solid) phase of the model. Y-axis is temperature, x-axis is average dimensionless density difference from a reference density (defined in the text).}
\label{fig_alPhaseDiag}
\end{figure}

The parameters $p_l$ and $q_l$ are chosen such that the model exhibits a thermodynamic phase diagram approaching that of pure aluminum in the vicinity of the triple point \cite{jreidini2022}. Figure \ref{fig_alPhaseDiag} compares the PFC model's temperature-density phase diagram with that of pure aluminum. The temperature $T$ is in units of Kelvin, while the average density $n_o$ is a rescaled dimensionless quantity related to real density through $n_o = (\rho-\bar{\rho})/\bar{\rho}$ where $\rho$ is the density  and  $\bar{\rho}$ is a reference density. The reference density was set near the solid coexistence line at a temperature appropriate for rapid quench experiments. The model's phase diagram exhibits a sharp deviation from pure aluminum's at densities below the critical point, due to the expanded nature of this model's free energy functional \citep{elder2007}. Further work is underway to expand PFC approaches to more accurately span the range of densities accessible in pure materials' phase diagrams \cite{kocher2019}. As the present work focuses on solidification, accurate agreement with the solidus in the vicinity of the triple point was determined to be sufficient.

Despite the proposed PFC model's diffusive-time evolution capabilities, it is still limited in spatial scales it can access due to the need to resolve individual atomic density peaks. To achieve larger simulation system scales, the model is transformed into a complex order parameter phase-field model, which does not resolve individual atomic peaks, through the use of amplitude expansions\citep{oforiopoku2013,athreya2006}. We expand the atomic density field as $n(\mathbf{x}) = \bar{n}(\mathbf{x}) + \sum_{j=1}^3 A_j(\mathbf{x})e^{i\mathbf{G}_j\cdot\mathbf{x}} + c.c.$ where $\bar{n}(\mathbf{x})$ is the long-wavelength average density field that does not include individual atomic peaks, $\mathbf{G}_j$ are the three lowest-order reciprocal lattice vectors for a triangular lattice, and $A_j(\mathbf{x})$ are the three complex order fields corresponding to each $\mathbf{G}_j$. After applying this expansion to equation \ref{eq_freeEnergy} and following volume-averaging techniques \citep{oforiopoku2013} we obtain an amplitude model defined by a free energy functional $F_{A}[A_j,\bar{n}]$ in terms of the coarse-grained fields. The evolution equations are obtained by assuming non-conserved dissipative dynamics for the complex order parameter fields $A_j(\mathbf{x})$ and conserved dissipative dynamics for the average density field $\bar{n}(\mathbf{x})$, i.e,
\begin{equation}\label{eq_evolutionAmp}
\frac{\partial A_j}{\partial t} = -\frac{\delta F_{A}}{\delta A_j^*} \quad \text{and} \quad \frac{\partial \bar{n}}{\partial t} = \nabla^2\frac{\delta F_{A}}{\delta \bar{n}}.
\end{equation}

The above amplitude field model was simulated in a comoving reference frame with an initial condition consisting of grains of random orientation up to $10\degree$ from a reference orientation.
Periodic boundaries were used for system edges perpendicular to the solidifying front. The average temperature of the system was set to $T=\SI{805}{\kelvin}$, a significant quench below pure aluminum's melting point of approximately $\SI{933}{\kelvin}$. 
Average density was chosen close to the solidus density.
A frozen temperature gradient of $G=\SI{5e7}{\kelvin\per\meter}$ was applied along the solid-liquid interface. 
The width and length of the simulation were approximately $\SI{400}{\nano\meter}$ and $\SI{2000}{\nano\meter}$ respectively. 
These simulation scales where chosen to be reasonably accessible numerically while still capturing the pertinent nanoscale phenomena emerging from rapid solidification. We expect essential simulated features discussed subsequently to be generically comparable to experimental results despite the difference in lengthscales.

\begin{figure*}
\centering
\includegraphics[width=\textwidth]{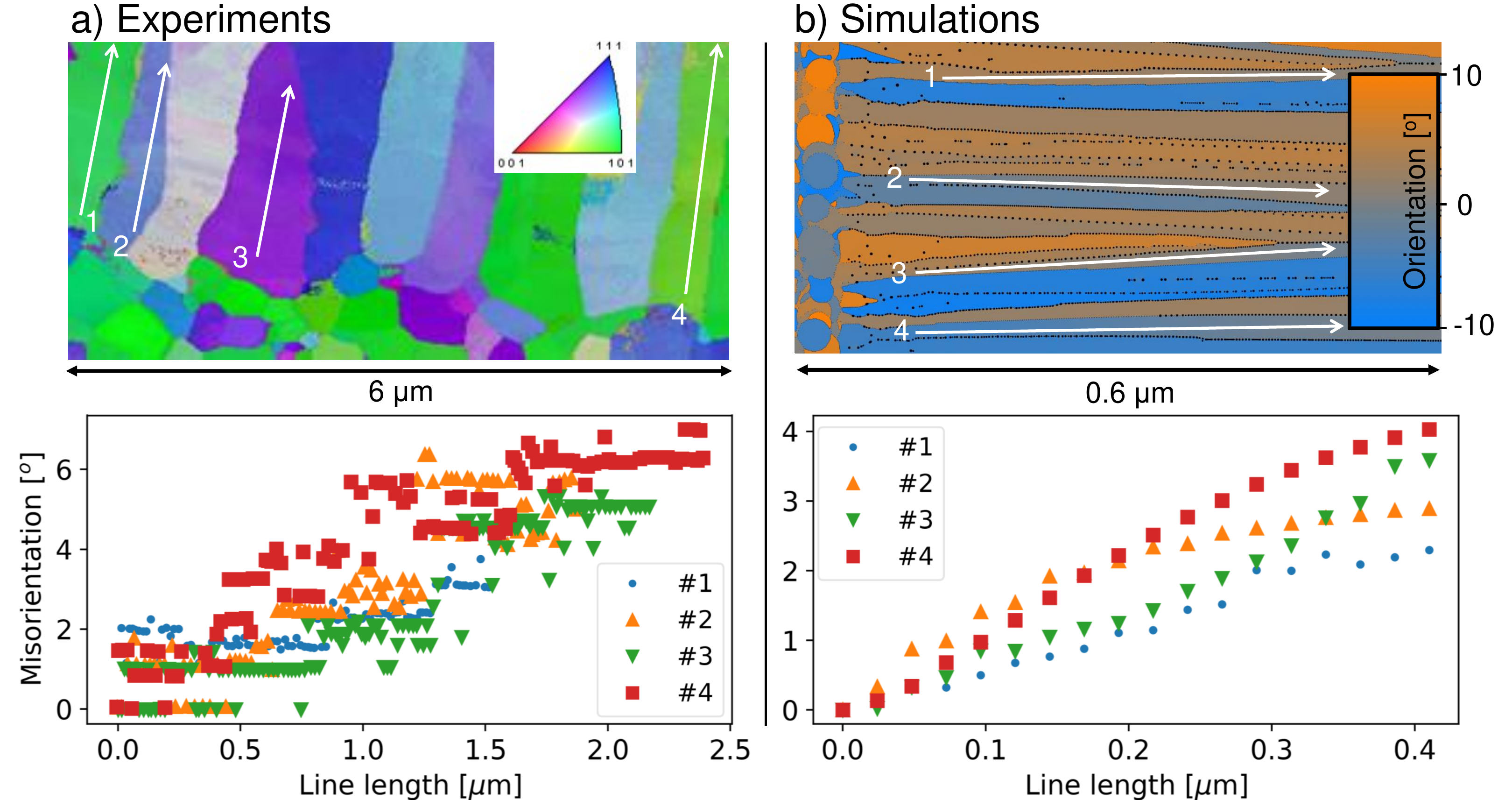}
\caption{Orientation distribution and plotted misorientation along labeled lines in a) laser thin film experiment and b) phase-field crystal simulation. Color map in top images is the same as in Figure \ref{fig_orientationGradients}. }
\label{fig_lineProfiles}
\end{figure*}

To compare with experimentally observed orientation gradients, the lattice orientations in the simulated solid grains were calculated from the atomic displacement field extracted from the complex amplitude fields $A_j$ \citep{heinonen2016}.
Figure \ref{fig_orientationGradients}b shows a collage of lattice orientations in a post-solidified simulated sample. These are reconstructed from snapshots of the comoving frame as it advanced into the melt, and are compared to similar microstructure observed in experiments (Figure \ref{fig_orientationGradients}a). We observe orientation gradients in the solid grains, in some cases on the order of $10\degree$ over $\SI{1000}{\nano\meter}$. Figure \ref{fig_lineProfiles} shows zoomed-in sections of the experimental system and the simulated Al system, plotting the change in orientation in individual grains as a function of distance. The simulated and experimental orientation gradient magnitudes (change in orientation per unit distance) are in fair agreement, with the simulations predicting a slightly larger average orientation gradient.

\begin{figure*}
     \centering
     \includegraphics[width=\textwidth]{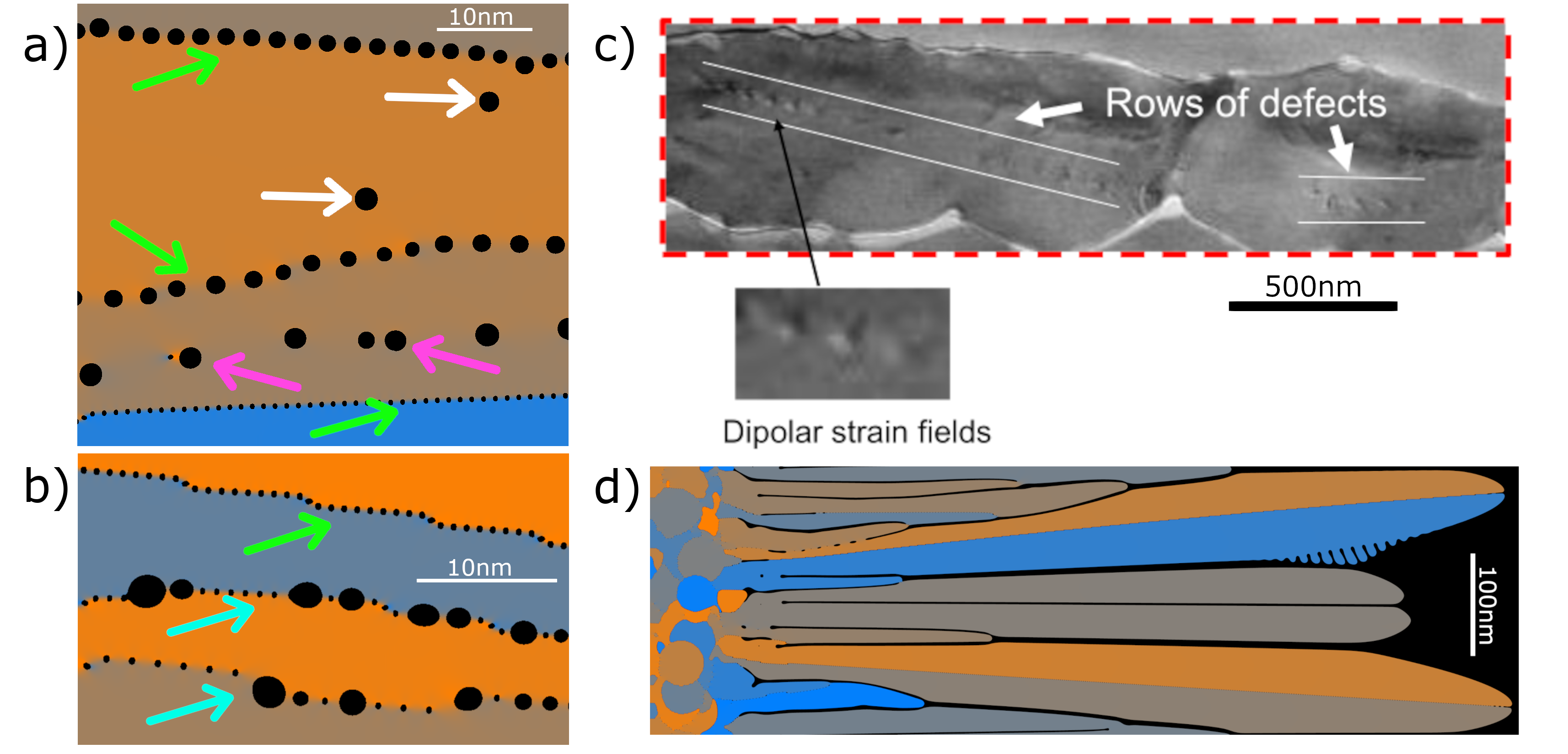}
        \caption{a) and b) A range of defect types observed in zoomed-in subsections of the simulated rapidly-solidified solid. White arrows show isolated voids within a single grain. Green arrows show rows of dislocations, each separating two grains of differing orientations. Though all these marked rows exhibit regular spacing and dipolar strain fields, the extent of ordering of the dislocation cores differs between rows.
        Further, the green-marked dislocation row in b) exhibits step-like structure, likely a result of lattice misfit accommodation \citep{furuhara1991,fujiwara2014,zhang2019}. Pink arrows show a row of what appears to be defects of an intermediate nature: unevenly spaced and sized, some exhibiting a dipolar strain field, and with inconsistent internal density. These emerge within a single grain that splits into two slightly different orientations as the solidification proceeds. Cyan arrows show defect arrays that consist of regularly spaced dislocations and voids, observed to interchangeably evolve diffusely in time as mentioned in the text.
        c) Rows of defects observed in a TEM image of an experimental sample. d) Collage of simulated polycrystalline  solidification of Al at a low cooling rate corresponding to a quench temperature of  $T=\SI{875}{\kelvin}$. The color map is the same as in Figure \ref{fig_orientationGradients}b.
        }
        \label{fig_lastcombined}
\end{figure*}

A closer look at the simulated solid bulks shown in Figures \ref{fig_orientationGradients}b and \ref{fig_lineProfiles}b reveals a variety of defect types. In the  simulation results displayed, the larger black areas in and between solid grains, where black corresponds to areas of low absolute value of the fields $A_j$, are found to be at densities approaching that of the vapor phase, while the smaller such black areas tend to be closer in density to the solid. A distinction can be made between these different sized defects: The larger, low-density areas are voids  (vapor pockets) formed due to density trapping at high interface velocities, while the smaller defects are dislocations. However, this distinction is not perfectly clear-cut, as many of the voids are found to exhibit dipolar strain fields indicating a topological nature equivalent to dislocations. Additionally, close observation of the system's time evolution reveals that some voids, particularly those along grain boundaries, can shrink into dislocations as density diffuses, and vice versa. A plausible cause for this interchangeability is the point-like nature of dislocations in two-dimensional systems, in contrast to their line-like nature in three-dimensional systems. Alternatively, this could be explained by the diffusive merger of dislocations to form voids. While such mergers might be a by-product of the coarse-grained nature of the PFC model and its derivatives, previous literature suggests that nanoscale defects can merge to form larger voids \cite{cuitino1996}. Figures \ref{fig_lastcombined}a and \ref{fig_lastcombined}b showcase simulated defect structures spanning this range of behaviors. Though experimental results tended to be of too low resolution to explicitly observe dislocations, the `mislabeled pixels' seen in the S/TEM inverse polar figure map of Figure \ref{fig_orientationGradients}a are likely due to crystalline defects. Further, some TEM images (shown in Figure \ref{fig_lastcombined}c) displayed what appeared to be rows of large-scale defects with dipolar strain fields, which are likely voids due to their relatively large interspacing.

During the dynamical simulations of rapid solidification, lattice orientation in solid grains is observed to evolve only within a few diffusion lengths of the solidifying front. Further behind this interface, the orientation appears `locked in' until the relevant part of the grain moves out of the comoving frame's limit.
The formation of orientation gradients is always accompanied by specific dislocation behavior at the solid-liquid interface. Namely, when a growing grain is observed to undergo a gradual rotation, the spacing of observed  dislocations forming at the boundary between this rotating grain and its neighbors is found to vary steadily as the rotation proceeds. Certain grains exhibit more chaotic behavior, splitting into multiple growing tips of slightly different orientations instead of rotating as a single lattice. This can be compared to the experimental images of Figure \ref{fig_orientationGradients}a, where such splitting is manifested as `sub-boundaries'. Again, this simulated behavior is observed to be closely tied to the formation of dislocations at the interface, often being preceded by sporadic trapped voids that exhibit a weak dipolar strain field surrounding them.  These observations collectively indicate that the formation of orientation gradients within the bulk is linked to dislocations and voids developing at the solidifying front.

To better illustrate the relation between defect formation and orientation gradients that emerge at rapid solidification rates, we also conducted simulations at a higher temperature quench $T=\SI{875}{\kelvin}$, in the limit of no frozen temperature gradient, i.e., $G=0$. The lower driving forces for solidification at these parameters leads to lower interface velocities.
Figure \ref{fig_lastcombined}d shows a collage of the total solidified material's lattice orientations. Unlike the case for rapidly solidified Al discussed above, here the inter-grain boundaries tended to remain `wet', consisting of metastable fluid, with fewer cases of `dry' boundaries containing dislocation arrays. The grains also displayed a propensity for dendrite arm formation, as expected for the lower driving forces involved. We observed in this and other simulations at low solidification rates that orientation gradients in solidified grains were much weaker and less common than in the case of the rapid solidification results reported above. Further, at low solidification rates, the only noticeable orientation gradients were observed in grains that shared a `dry' grain boundary. Our results thus suggest that the appearance of orientation gradients is a phenomenon intrinsically tied to the high interface velocities found in rapid solidification regimes, and again support the hypothesis that the experimentally-observed orientation gradients are connected to the formation of dislocation arrays at grain boundaries.

In summary, this work reports new observations of  orientation gradients in pure aluminum thin film resolidification experiments as well as in simulations conducted with a novel 3-phase single-component PFC model of aluminum. The simulated orientation gradient magnitudes are in good qualitative agreement with experiments.
The likely presence of defects in the experimental images, as indicated by mislabeled pixels in the inverse polar figure maps as well as the observed presence of defect arrays, agrees with simulation results and suggests the importance of dislocations and voids in mediating the formation of orientation gradients within single grains. Future work on this topic would benefit from higher resolution experimental imaging techniques capable of examining the discussed  defects at lengthscales comparable to that of the simulations. Also of interest would be attempting simulations with the developed PFC model in three dimensions to better capture defect topologies, and extending the model to include a dynamically co-evolving temperature field \cite{kocher2019}.

%%%%%%%%%%%%%%%%%%%%%%%%%%%%%%%%%%%%%%%%%%%%%%%%%%%%%%%%%%%%%%%%%%%%
%%%%%%%%%%%%%%%%%%%%%%%%%%%%%%%%%%%%%%%%%%%%%%%%%%%%%%%%%%%%%%%%%%%%
%%%%%%%%%%%%%%%%%%%%%%%%%%%%%%%%%%%%%%%%%%%%%%%%%%%%%%%%%%%%%%%%%%%%

%\appendix
\begin{acknowledgments}
Work at McGill university received support from the National Science and Engineering Research Council of Canada (NSERC) and the Canada Research Chairs (CRC) Program.
Work at VTT was supported by Academy of Finland through the HEADFORE project, Grant No. 333226.
Work at the University of Pittsburgh received support from the National Science Foundation under grant number DMR 1607922. Work at Lawerence Livermore National Laboratory was performed under the auspices of the U.S. Department of Energy under Contract DE-AC52-07NA27344.
\end{acknowledgments}

%\section{Appendixes}

%\bibliography{apssamp}% Produces the bibliography via BibTeX.

\bibliographystyle{unsrt}
\bibliography{bibliography}

\end{document}